\documentclass[sn-mathphys,Numbered]{sn-jnl}

\usepackage{graphicx}%
\usepackage{multirow}%
\usepackage{amsmath,amssymb,amsfonts}%
\usepackage{amsthm}%
\usepackage{mathrsfs}%
\usepackage[title]{appendix}%
\usepackage{xcolor}%
\usepackage{textcomp}%
\usepackage{manyfoot}%
\usepackage{booktabs}%
\usepackage{algorithm}%
\usepackage{algorithmicx}%
\usepackage{algpseudocode}%
\usepackage{listings}%
\usepackage{subcaption}%

\definecolor{Rev}{rgb}{0,0,0}

\theoremstyle{thmstyleone}%
\theoremstyle{thmstyletwo}%

\theoremstyle{thmstylethree}%

\begin{document}

\title{A finite element approach for modelling the fracture behaviour of unidirectional FFF-printed parts}


\author*[1]{\fnm{Simon} \sur{Seibel}}\email{simon.seibel@unibw.de}

\author[1]{\fnm{Josef} \sur{Kiendl}}\email{josef.kiendl@unibw.de}

\affil*[1]{\orgname{University of the Bundeswehr Munich, Institute of Engineering Mechanics \& Structural Analysis}, \orgaddress{\street{Werner-Heisenberg-Weg 39}, \city{Neubiberg}, \postcode{85577},  \country{Germany}}} 

\abstract{We present a finite element modelling approach for unidirectional Fused Filament Fabrication (FFF)-printed specimens under tensile loading. In this study, the focus is on the fracture behaviour, the goal is to simulate the mechanical behaviour of specimens with different strand orientations until final failure of the specimens. In particular, the aim is to represent experimentally observed failure modes for different print orientations and the typical dependence of the parts' strength on the print orientation. We investigate several modelling aspects like the choice of a suitable failure criterion, a suitable way to represent fracture in the finite element mesh or the necessary level of detail when modelling the characteristic edges of FFF-printed specimens. \textcolor{Rev}{As a result, this work provides an approach to model FFF printed specimens in finite element simulations, which can represent the characteristic relation between mesostructural layout and macroscopic fracture behaviour.}}




\maketitle
\newpage

\section{Introduction}\label{Introduction}

Additive manufacturing (AM) has found its way into rapid prototyping and the production of small quantities like spare parts in particular. After creating a geometry in a CAD software, a 3D surface is generated. This surface is processed by a so-called slicer software into a G-code, which contains all the commands for the printing process. There are various techniques to print the parts, including Fused Filament Fabrication (FFF), which is the focus of the following studies.
In FFF printing, the mechanical properties of the printed part depend \textcolor{Rev}{without consideration of lattice structures at the macro level, as shown, for example, in \cite{Khosravani.2024}}, on many parameters, such as the printing speed, the printing material, the bed temperature, the printing temperature, \textcolor{Rev}{ageing,} the type of retraction and the spacing of the strands (\cite{Djokikj.2022},\cite{Butt.2022},\cite{Popescu.2018},\cite{Yan.2022},\cite{Petousis.2023},\textcolor{Rev}{\cite{Khosravani.2020}}). In particular, the mechanical properties in FFF depend on the layer orientation, which numerous studies (\cite{Kiendl.2020},\cite{Khosravani.2022},\cite{Zhang.2023},\cite{Song.2017},\cite{Yao.2020},\cite{Alaimo.2017},\cite{Li.2002}) have shown experimentally. The main result of these studies with a unidirectional layer structure of the tensile specimens states that e.g. the tensile strength is highest if the strand direction is equal to the loading direction and lowest if the strand direction is orthogonal to the loading direction. This effect, that the tensile strength and stiffness decrease with increasing angle between strand orientation and loading direction, was also shown partly in FE simulations, that can be divided in two methods. In principle, there are methods of homogenised models, for example using a representative volume element (RVE) or classical laminate theory (CLT) theory (\cite{Gonabadi.2022},\cite{Somireddy.2018}) or models, that resolve the micro-/mesostructure (\cite{Voelkl.2020},\cite{Wendt.2017}) to predict the mechanical properties. Models using a RVE or CLT can model the stiffness and failure as a function of different angles by determining the material parameters for each configuration from several reference tests and then simulating (parts of) the tensile specimen.  In contrast to the models using anisotropic material parameters, the number of material parameters as input for models can be reduced by resolving the mesostructure. Only the structure of the strands with the same (isotropic) material parameters for every configuration results in different mechanical properties on a macroscopic level. \cite{Garg.2017}, for example, simulates the complex behaviour of the mesostructure using an elasto-plastic approach, but the failure is not simulated. The number of homogenised models outnumbers models that resolve the individual strands and failure of the specimen itself was simulated in none of the mentioned studies. This is where this work starts and simulating the failure in the Finite Element Analysis (FEA) to show the dependency of the angle between strands and the loading direction and simulation of the failure on a mesostructure scale. The advantage of modelling the specimens using individual strands is the outcome of realistic failure patterns and the lower input of material parameters, as the effect of the different orientations is provided by the strand's geometry and orientation in relation to the loading direction.
Therefore, the results of \cite{Kiendl.2020} are used in this study, as not only the angles $0^{\circ}$, $45^{\circ}$ and $90^{\circ}$ were experimentally investigated here, but also angles in between. The aim is to numerically simulate the mechanical behaviour of unidirectional FFF-printed specimens with different print orientations until final failure, representing the experimentally observed failure modes and the typical dependence of the parts' strength on the print orientation.

\section{Materials and Methods}

\subsection{FFF printing}\label{FFF printing}
\begin{figure}[h]
	\centering
	
	\begin{subfigure}{0.4\textwidth}
		\includegraphics[width=\textwidth]{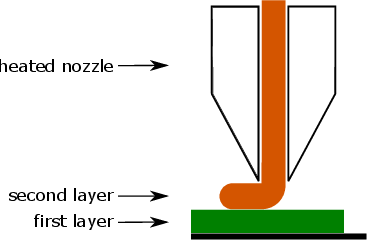}
		\caption{}
	\end{subfigure}
	\hfill
	\begin{subfigure}{0.4\textwidth}
		\includegraphics[width=\textwidth]{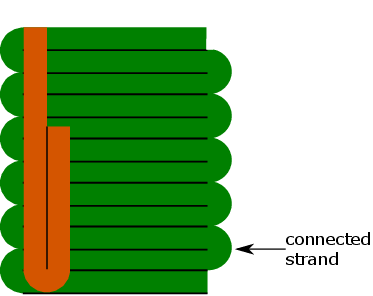}
		\caption{}
	\end{subfigure}
	
	\caption{FFF printing process (a) from the side view and (b) from the top view (without the nozzle) }\label{FFF_process}
\end{figure}

In this section, the basic features of the FFF printing process and the resulting mesostructure, which are shown schematically in figure \ref{FFF_process}, and its impact on the mechanical material behaviour of the printed part are discussed. 

\begin{figure}[h]
	\centering
	\includegraphics[width=0.5\textwidth]{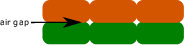}
	\caption{Schematic section through the printed object with two layers printed in the same direction}\label{air_gaps}
\end{figure}

In the FFF process, a thermoplastic filament is melted from a coil through a heated nozzle and then deposited on a print bed along a predefined path. This procedure is repeated layer by layer until the object is fully printed. Typically, a layer is printed continuously so that the (mostly parallel) strands are always connected at one end.  This could have an influence on the type and the initiation of failure. Due to the dependency of the mechanical properties on the print orientation, the layers are typically printed alternately at a $90^{\circ}$ angle in order to minimise these dependencies and to ensure that the failure does not depend too much on the print orientation. In this work we want to focus on precisely this resulting anisotropy and print the tensile specimens with a unidirectional layup.
Typical for the printing process are the air gaps between the strands shown in figure \ref{air_gaps}. The strands touch each other on all sides, but not over the entire width or height due to the air gaps, which also has an influence on the failure.

\begin{figure}[H]
	\centering
	\includegraphics[width=0.5\textwidth]{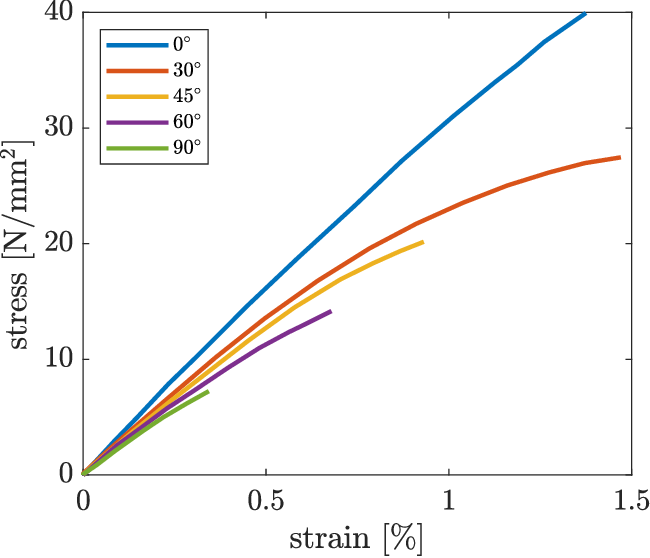}
	\caption{Experimental results of unidirectional layup \cite{Kiendl.2020}}\label{experiments_chao}
\end{figure}

The direction of the strands can be varied as required for one and the same object. This resulting anisotropy, which comes purely from the printing process, has a considerable influence on the mechanical properties of the object as mentioned above. As an example, in addition to the studies mentioned in section \ref{Introduction}, figure \ref{experiments_chao} shows the stress-strain-diagram for a tensile test with different angles of the deposited strands relative to the tensile direction. The decreasing strength and stiffness with increasing angle are clearly recognisable. \textcolor{Rev}{The specimens in \cite{Kiendl.2020} were printed on a Raise3D Pro2 printer with a extruder width of $0.4$ $mm$ and PLA filament from EasyPrint was used. Printing speed was set to $30$ $\frac{mm}{s}$. No contour were used, which would have a potential influence on the layers. The extruder temperature was set to $200^{\circ}C$, the bed temperature to $60^{\circ}C$. For further details, reference is made to \cite{Kiendl.2020}.} The experiments from \cite{Kiendl.2020}, shown in figure \ref{experiments_chao}, are compared and discussed with the numerical model presented in this paper.

Additionally, the polymer strand can consist of different materials such as acrylonitrile butadiene styrene (ABS), polyether ether ketone (PEEK) or polylactic acid (PLA), which means different mechanical properties for the printed object at the end. In this work, the focus is on PLA.

\subsection{Numerical model}\label{implementation}
\begin{figure}[h]
	\centering
	\includegraphics[width=0.4\textwidth]{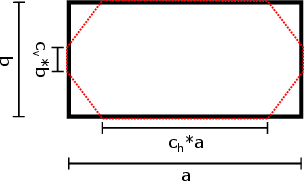}
	\caption{Approximation of strand's geometry}\label{approx_geom}
\end{figure}

In this section the modelling assumptions and numerical setup for the simulations are explained. Ansys\textsuperscript{\copyright} Mechanical Enterprise, Release 22.2 \cite{AnsysGeneral.2022} is the commercial FE software that was used for the simulations. All simulations were performed on a Linux HPC cluster using parallel computing with up to 192 processors. All challenges are briefly explained and a possible approach to the problem is outlined. \\
To resolve the mesostructure of the tensile specimen each strand is modelled individually and an octagon cross-section is used to approximate the oval shape of the printed strand as shown in figure \ref{approx_geom}. 

To determine the factors $c_v$ and $c_h$, that are multiplied with the width $a$ respectively height $b$, a microscope image of a cross-section of a specimen after failure is used and the factors are measured and averaged over several strands. Details will be explained in the section \ref{geometry}.\\
Experimental tests with unidirectional PLA specimens have shown a rather brittle material behaviour, as can be seen in the stress-strain diagram in figure \ref{experiments_chao} and in the fracture surface shown in figure \ref{fracture_image}. Accordingly, linear isotropic elastic material behaviour with a brittle failure is assumed. To describe this material behaviour, only the three material parameters Young's modulus $E$, Poisson's ratio $\nu$ and the strength $\sigma_{f}$ are required. These are determined using reference tests in section \ref{material model}.\\
Since a large number of elements are required for a detailed resolving of the mesostructure, only one layer and only a part of the specimen are simulated. This is described in more detail in section \ref{Mesh}. For the fracture, the techniques of Element Erosion and Element Deletion are presented in section \ref{EE} and discussed in section \ref{EEvsED}. These are simulation techniques in which the stiffness of an element is greatly reduced in the case of Element Erosion or the entire element is deleted in the case of Element Deletion once a user-defined criterion has been reached. Different equivalent stresses (von Mises equivalent stress and maximum-normal stress), which calculate the criterion for the failure, are presented in section \ref{material model} and discussed in section \ref{unidirectional}. In addition, case studies on the different designs and their influences of the specimen's edges are performed. It is possible to model the connected strands true to detail, to model the strands slightly offset to each other or the influence of the edge is negligible and no additional edge is modelled. The comparison is made in section \ref{edge}. \\

\subsubsection{Geometry}\label{geometry}
The strands in \cite{Kiendl.2020} were printed with a width $a=0.4$ $mm$ and a height of $b=0.2$ $mm$. 

\begin{figure}[h]
	\centering
	\includegraphics[width=0.4\textwidth]{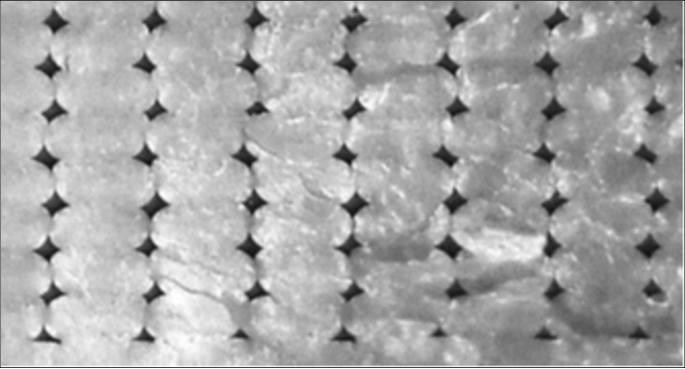}
	\caption{Fracture image of a $0^{\circ}$ tensile specimen}\label{fracture_image}
\end{figure}

If we look at the contact lengths of all the strands shown in figure \ref{fracture_image} and average them, we get on average a value of $c_h=0.72$ and $c_v=0.36$. While the scatter of the horizontal contact length is very small with a standard deviation of $\sigma = 0.032$, the value for the vertical contact length scatters much more with a standard deviation of $\sigma = 0.066$. Zones in which the strands from two different layers touch are called interlayer zones, whereas zones in which two strands from one layer touch are called intralayer zones. 
It has been shown experimentally e.g. in \cite{Kiendl.2020}, \cite{Khosravani.2022} and \cite{Yao.2020} that unidirectional specimen typically fail due to intralayer debonding, unless loading is parallel to the strand orientation.

Figure \ref{approx_geom} shows the approximated shape of the strand. \textcolor{Rev}{Every strand is resolved uniformly and small differences of different strands resulting from the printing process are neglected. The contact length $c_v$ has an high influence on the failure and on the material response on a macroscopic level e.g. the strength $\sigma_{f}$, as explained later in \ref{intralayer_zone}. The most simplified form to capture these effects is the octagon to approximate the strand.} $a$ and $b$ correspond to the width and height of the printed specimen, that are set in the slicer for 3D printing. It is known that the material strength in the intralayer zones is generally lower than in the bulk of the strands due to limited molecular diffusion between strands during printing \cite{Li.2018,Balani.2023,Coogan.2020}. 
For taking into account, the strength is reduced in the intralayer zones, the vertical contact length $c_v$, which was determined optically in section \ref{geometry} is reduced, as shown in detail in section \ref{intralayer_zone}.

\subsubsection{Material model}\label{material model}
For a linear elastic material model with brittle failure, three material parameters are required. While the Poisson's ratio $\nu$ is taken from the literature as $\nu = 0.3$ like in \cite{Mirkhalaf.2021}, the Young's modulus $E$ and the strength $\sigma_{f}$ are taken from the experiments from \cite{Kiendl.2020} shown in figure \ref{experiments_chao}, more precisely, from the $0^{\circ}$ case. 
It has to be noted that in \cite{Kiendl.2020}, stresses were calculated from the measured force assuming a solid rectangular cross-section of the specimen (ignoring the voids).
Therefore, stiffness and strength for our model with the octagon cross-section of the strands (see figure \ref{approx_geom}) are obtained by scaling the values extracted from figure \ref{experiments_chao} by the ratio 

\begin{align}
	\frac{A_{rec}}{A_{oct}},
\end{align}

where $A_{rec}=a\cdot b$ is the cross-sectional area of a simplified, rectangular shape of a strand, and $A_{oct}$ is the cross-sectional area of the octagon shape in figure \ref{approx_geom}. With $c_h=0.7$ and $c_v=0.2$, that are initially chosen under the assumptions (reduced strength in the intralayer zones) from section \ref{geometry}, we obtain

\begin{align}
	\frac{A_{rec}}{A_{oct}}= 1.136.
\end{align}

Accordingly, the Young's modulus and strength for the simulation models are obtained as 

\begin{align}
	E_{sim} &= 1.136 \cdot E_{exp} \\
	\sigma_{f,sim} &= 1.136 \cdot \sigma_{f,exp},
\end{align}

where $E_{exp}$ and $\sigma_{f,exp}$ are the values obtained from the $0^{\circ}$ specimen experimental curve in figure \ref{experiments_chao}. 

All material parameters for all upcoming simulations are summarized in table \ref{material_parameter}.

\begin{table}
	\begin{tabular}{c  c  c}
		\toprule
		$E_{sim}$		& $\nu_{sim}$	& $\sigma_{f,sim}$ \\
		$3901$ $MPa$	& $0.3$ & $45.38$ $MPa$\\
		\bottomrule
	\end{tabular}
	\caption{Material parameters}\label{material_parameter}
\end{table}

Since 3D solid elements are used in the simulations, an equivalent stress measure is needed for the failure criterion. Two different equivalent stresses are used and compared for the subsequent numerical implementation.  Equivalent stresses can be calculated with the help of the principal stresses $\sigma_I$,  $\sigma_{II}$ and  $\sigma_{III}$. Besides the von Mises equivalent stress

\begin{equation}\label{vonMises}
	\sigma_{eqv,vM}=\sqrt{\frac{(\sigma_I-\sigma_{II})^2+(\sigma_{II}-\sigma_{III})^2+(\sigma_{III}-\sigma_I)^2}{2}},
\end{equation} 

which is particularly suitable for ductile materials \cite{Mises.1913}, the maximum-normal stress hypothesis is used for brittle materials and/or rapid loading \cite{Hearn.1997}\textcolor{Rev}{,\cite{Sackey.2022},\cite{Gu.2018}}. The maximum-normal stress hypothesis also known as Rankine's theory \cite{Rankine.1857} calculates the equivalent stress for tension-dominated stress states ($\sigma_I \geq 0$ and $\sigma_I \geq |\sigma_{III}|$) as

\begin{equation}\label{Rankine}
	\sigma_{eqv,R}=\sigma_I
\end{equation} 

as a criterion for failure.

\subsubsection{Simulation of the fracture}\label{EE}
\textcolor{Rev}{To simulate failure Ansys\textsuperscript{\copyright} offers the Element Birth \& Death technology, also known as Element Erosion (EE) \cite{AnsysEE.2022}. Since this technique can lead to convergence problems (see chapter \ref{EEvsED}), we developed an alternative approach in which failure is represented with deleted elements instead of eroded elements. For this purpose, data like element numbers, material parameters, previous computed results etc. must be temporarily stored after each load step/iteration (if failure happens) and transferred to a new mesh. The next load step/iteration is then computed with a new mesh. We call this technique Element Deletion (ED). The principle of these two methods are explained in the following for pointing out the differences.}

After an initial simulation, each element is checked in post-processing for exceeding a freely selectable criterion. This criterion can be e.g equivalent stresses, strains or any other user-defined criterion. If the element exceeds this limit, the element will be "killed" or deleted. It is important to know that "killing" the element does not mean that the element is completely deleted, but the element stiffness is multiplied with the factor $10^{-6}$. Thus the element still exists in contrast to ED, where the whole element is completely deleted.\\
The advantages of EE are:

\begin{itemize}
	\item continuos connectivity of the FE mesh
	\item visualization of the failed elements.
\end{itemize}

Whereas the disadvantages for EE are:

\begin{itemize}
	\item excessive element distortion of eroded elements, although physically irrelevant, can lead to abortion of the simulation
\end{itemize}

The advantages and disadvantages for ED are vice versa. It is noted, that both techniques were used for the simulations and both lead almost to the same results and are shown in \ref{EEvsED}. 

\begin{figure}[h]
	\centering
	\includegraphics[width=1\textwidth]{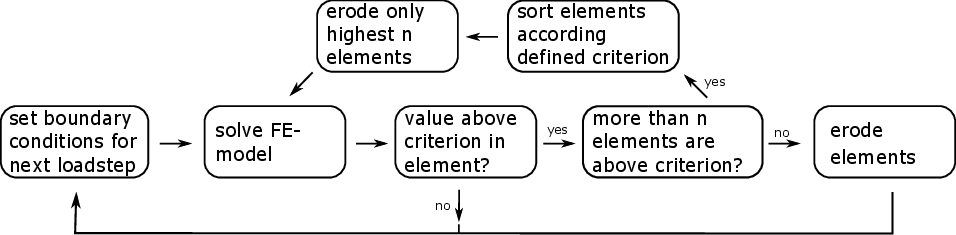}
	\caption{Extended scheme for an EE simulation with localization }\label{SchemeEE_localized}
\end{figure}

\begin{figure}[h]
	\centering
	\includegraphics[width=1\textwidth]{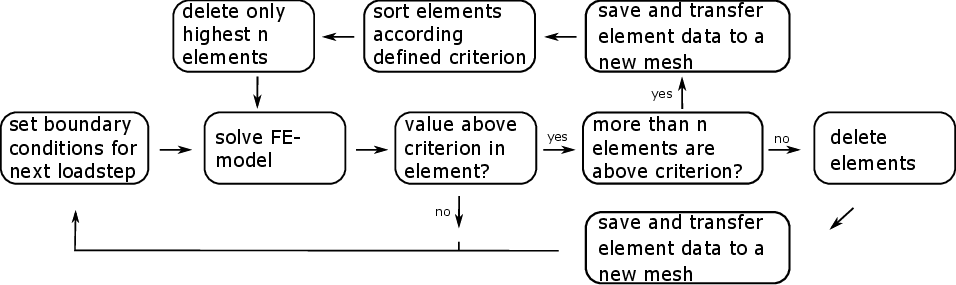}
	\caption{Extended scheme for an ED simulation with localization }\label{SchemeED_localized}
\end{figure}

Due to the uniform stress state in these simulations, many elements exceed the failure criterion at the same time. This occurs mainly at the intralayer zones. To prevent all intralayer zones from failing at the same time the failure must be localized. To implement this localization, the software's workflow is extended and shown in figure \ref{SchemeEE_localized} for EE and in figure \ref{SchemeED_localized} for ED.

If the element solution of at least one element exceeds the user-defined criterion, all elements that exceed this criterion are sorted in descending order. In order to achieve localization, not all elements may be "killed" or deleted, but only the $n$ elements, that exceed the criterion value the highest. The amount $n$ must be chosen sensibly so that neither too few elements are "killed" or deleted and the simulation becomes too expensive, nor too many elements are "killed" or deleted and no localization takes place. In the simulations in section \ref{results}, a value of $0.01$ $\%$ of the total number of elements was selected for $n$.

\subsubsection{Mesh}\label{Mesh}
\begin{figure}[h]
	\centering
	\includegraphics[width=0.2\textwidth]{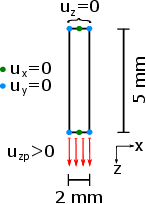}
	\caption{Simulated part of the specimen including boundary conditions}\label{simulation_area}
\end{figure}

The mesh for the FE simulations was created directly in the simulation software Ansys\textsuperscript{\copyright}. A 3-D 8-node solid element using the $\overline{B}$-method with full integration was used. \textcolor{Rev}{The converge criteria are $0.5$ $\%$ for displacements and forces to ensure a realistic and valid outcome.} Every node has three degrees of freedom, that are the translations in x, y and z direction. 
Our models do not simulate the whole dogbone from \cite{Kiendl.2020}, but only one layer and a smaller part shown in figure \ref{simulation_area} to reduce the amount of elements. At the upper and lower ends, the displacement in the x-direction is fixed in the centre and the displacement in the y-direction is fixed at the respective corners. In addition, the entire upper edge is fixed in the z-direction and the displacement is applied in every loadstep to the lower edge until the specimen fails. The left and right edge are traction free. The boundary conditions are shown in figure \ref{simulation_area}.

\begin{figure}[h]
	\centering
	
	\begin{subfigure}{0.4\textwidth}
		\includegraphics[width=\textwidth]{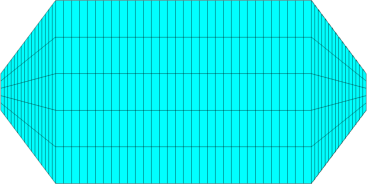}
		\caption{}
	\end{subfigure}
	\hfill
	\begin{subfigure}{0.4\textwidth}
		\includegraphics[width=\textwidth]{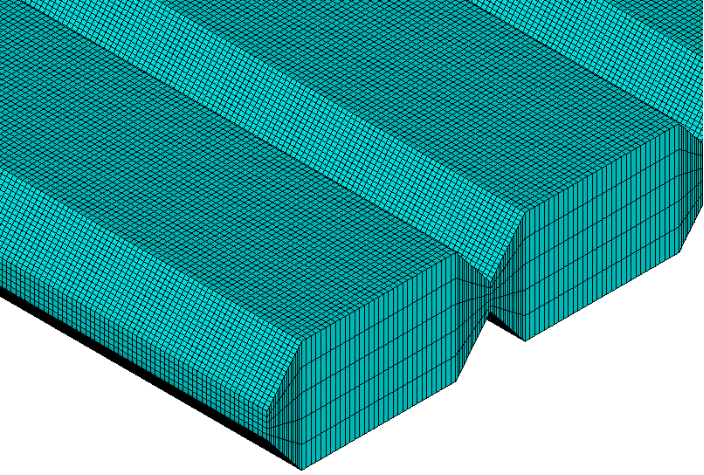}
		\caption{}
	\end{subfigure}
	
	\caption{(a) Mesh of a strand cross-section with 320 8-node hexaeder and (b) section of the entire mesh }\label{mesh_filament}
\end{figure}

Requirements for the mesh for the simulations of the mesostructure are, on the one hand, to limit the number of elements upwards so that the simulation does not become too expensive, but on the other hand, \textcolor{Rev}{to have the mesh fine enough for representing failure by narrow zones of eroded/deleted elements, avoiding too much "loss" of material.} The mesh of a cross-section and a section of the entire mesh are shown in figure \ref{mesh_filament}. The simulation models have an average total number of 500,000 elements. The exact number of elements depends on the angle of the strands to the loading direction. The average calculation time of a model on 96 processors on the cluster is approximately 24 hours. The central area of the dogbone in \cite{Kiendl.2020} has a dimension of $10$ $mm \cdot 50$ $mm$. To simulate the entire dogbone for our studies, the number of elements would be in the tens of millions, which would not be practical.

\section{Results and discussions}\label{results}
This section presents the results of the simulations and parameter studies. Firstly, the techniques EE and ED, the different stress hypotheses plus the influence of the specimen's edge are discussed in order to qualitatively build the model. Secondly, the model is then finally calibrated using the contact lengths $c_v$ and $c_h$. All experiments that are simulated are from \cite{Kiendl.2020}, that are shown in figure \ref{experiments_chao}.

\subsection{Element Erosion vs. Element Deletion}\label{EEvsED}

\begin{figure}[h] 
	\centering
	
	\begin{subfigure}{0.165\textwidth}
		\includegraphics[width=\textwidth]{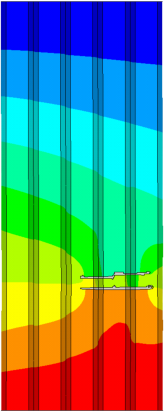}
		\caption{}
	\end{subfigure}
	\hfill
	\begin{subfigure}{0.1525\textwidth}
		\includegraphics[width=\textwidth]{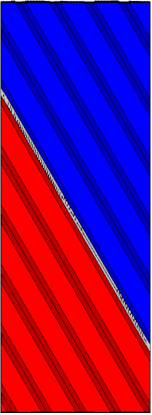}
		\caption{}
	\end{subfigure}
	\hfill
	\begin{subfigure}{0.157\textwidth}
		\includegraphics[width=\textwidth]{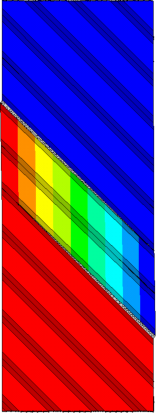}
		\caption{}
	\end{subfigure}
	\hfill
	\begin{subfigure}{0.155\textwidth}
		\includegraphics[width=\textwidth]{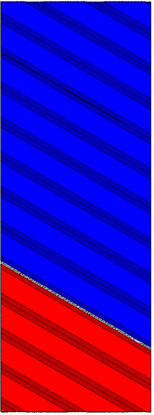}
		\caption{}
	\end{subfigure}
	\hfill
	\begin{subfigure}{0.1625\textwidth}
		\includegraphics[width=\textwidth]{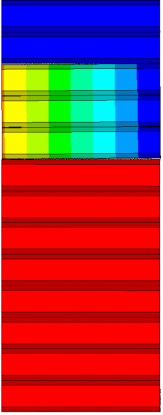}
		\caption{}
	\end{subfigure}
	
	\begin{subfigure}{1\textwidth}
		\includegraphics[width=\textwidth]{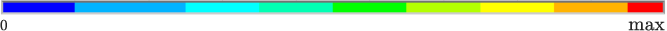}
		\caption{}
	\end{subfigure}
	
	\caption{Failure patterns for the simulations for (a) $0^{\circ}$, (b) $30^{\circ}$, (c) $45^{\circ}$, (d) $60^{\circ}$ and (e) $90^{\circ}$ for the EE-technique. The colorbar (f) shows the displacement in loading direction.} \label{Bruchbilder_EE}
	
\end{figure}

\begin{figure}[h] 
	\centering
	
	\begin{subfigure}{0.3\textwidth}
		\includegraphics[width=\textwidth]{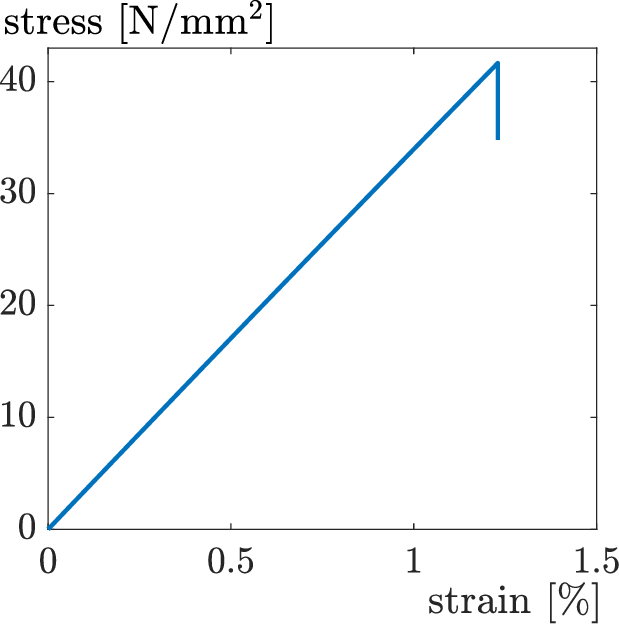}
		\caption{}
	\end{subfigure}
	\hfill
	\begin{subfigure}{0.3\textwidth}
		\includegraphics[width=\textwidth]{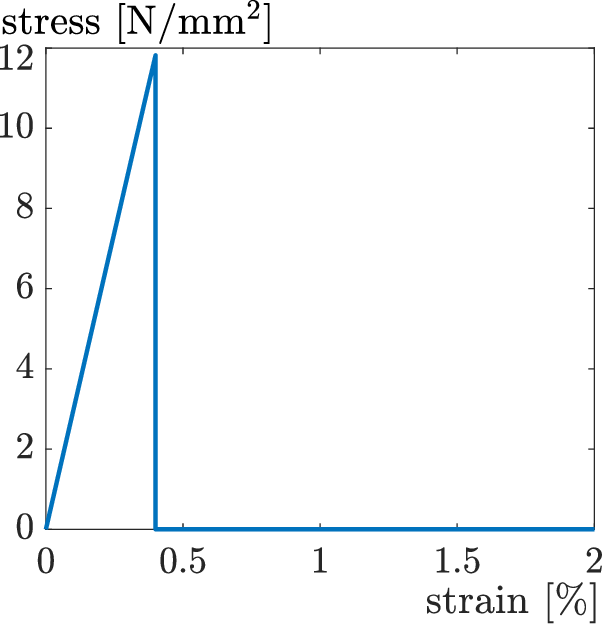}
		\caption{}
	\end{subfigure}
	\hfill
	\begin{subfigure}{0.3\textwidth}
		\includegraphics[width=\textwidth]{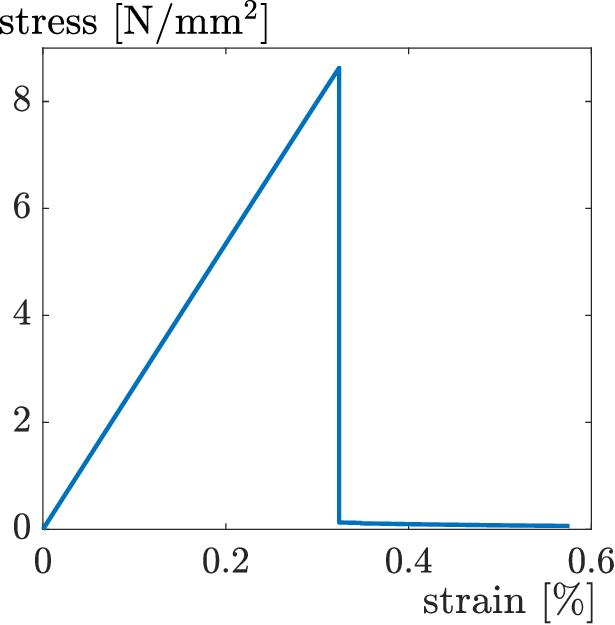}
		\caption{}
	\end{subfigure}
	
	\begin{subfigure}{0.3\textwidth}
		\includegraphics[width=\textwidth]{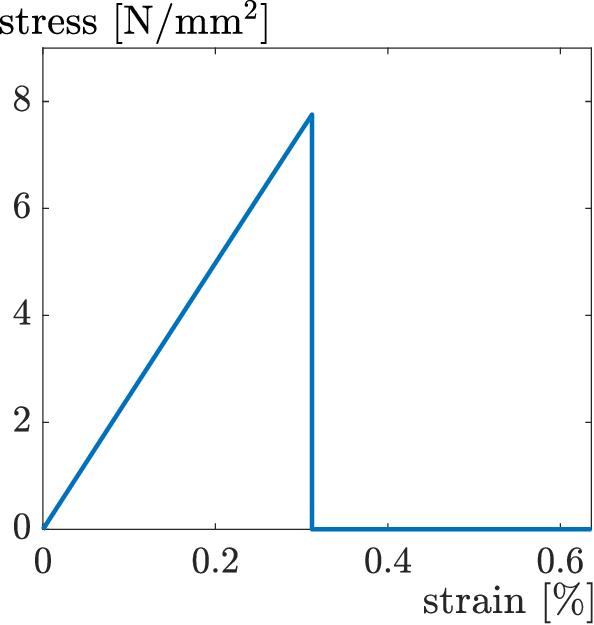}
		\caption{}
	\end{subfigure}
	\hfill
	\begin{subfigure}{0.3\textwidth}
		\includegraphics[width=\textwidth]{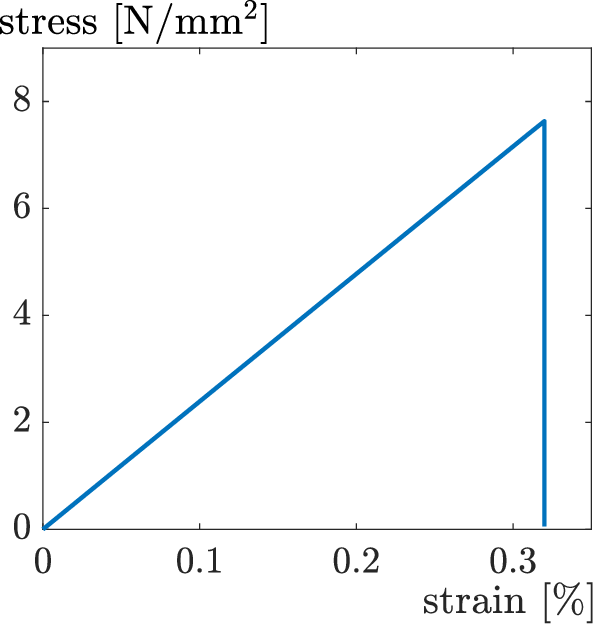}
		\caption{}
	\end{subfigure}
	
	\caption{stress-strain-diagrams for the simulations for (a) $0^{\circ}$, (b) $30^{\circ}$, (c) $45^{\circ}$, (d) $60^{\circ}$ and (e) $90^{\circ}$ using the EE-technique.} \label{Plot_EE}
	
\end{figure}

\begin{figure}[h] 
	\centering
	
	\begin{subfigure}{0.165\textwidth}
		\includegraphics[width=\textwidth]{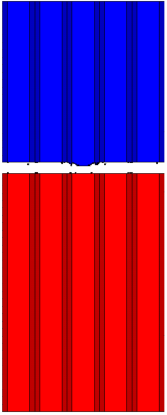}
		\caption{}
	\end{subfigure}
	\hfill
	\begin{subfigure}{0.151\textwidth}
		\includegraphics[width=\textwidth]{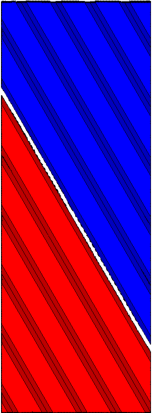}
		\caption{}
	\end{subfigure}
	\hfill
	\begin{subfigure}{0.1515\textwidth}
		\includegraphics[width=\textwidth]{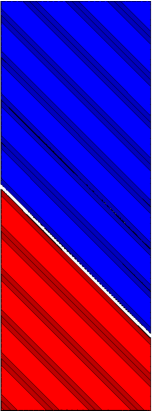}
		\caption{}
	\end{subfigure}
	\hfill
	\begin{subfigure}{0.15\textwidth}
		\includegraphics[width=\textwidth]{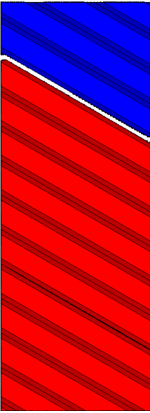}
		\caption{}
	\end{subfigure}
	\hfill
	\begin{subfigure}{0.16125\textwidth}
		\includegraphics[width=\textwidth]{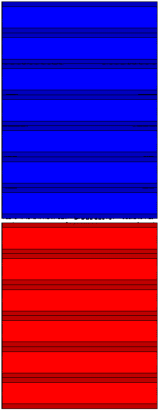}
		\caption{}
	\end{subfigure}
	
	\begin{subfigure}{1\textwidth}
		\includegraphics[width=\textwidth]{colorbar.eps}
		\caption{}
	\end{subfigure}
	
	\caption{Failure patterns for the simulations for (a) $0^{\circ}$, (b) $30^{\circ}$, (c) $45^{\circ}$, (d) $60^{\circ}$ and (e) $90^{\circ}$ for the ED-technique. The colorbar (f) shows the displacement in loading direction.} \label{Bruchbilder_ED}
	
\end{figure}

\begin{figure}[h] 
	\centering
	
	\begin{subfigure}{0.3\textwidth}
		\includegraphics[width=\textwidth]{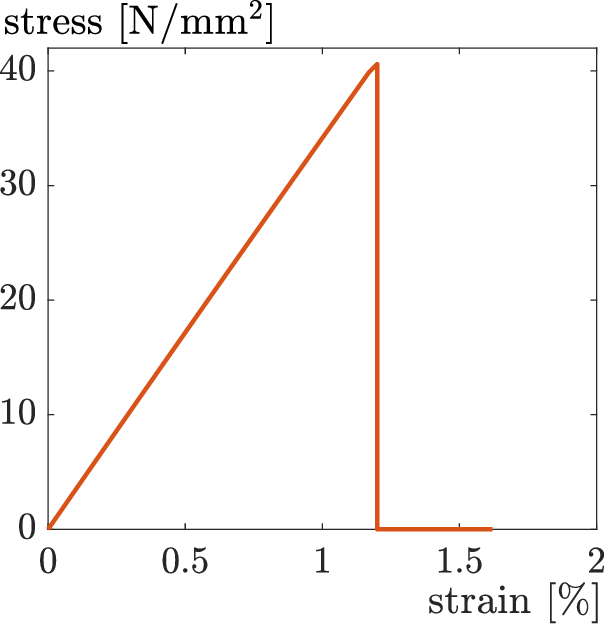}
		\caption{}
	\end{subfigure}
	\hfill
	\begin{subfigure}{0.3\textwidth}
		\includegraphics[width=\textwidth]{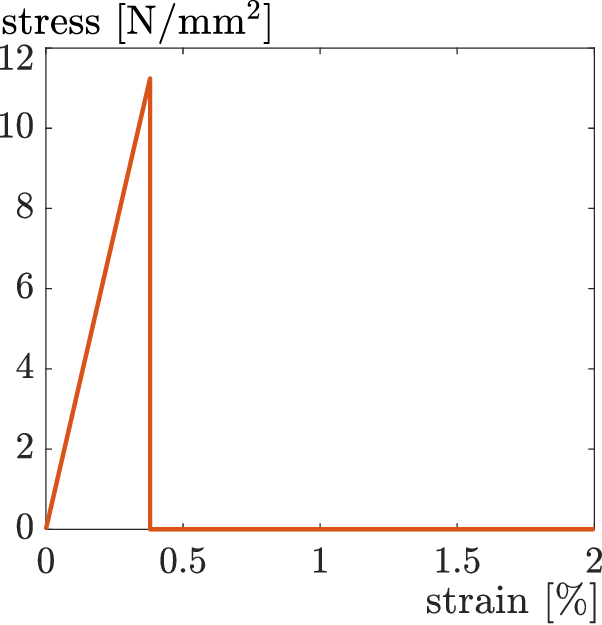}
		\caption{}
	\end{subfigure}
	\hfill
	\begin{subfigure}{0.3\textwidth}
		\includegraphics[width=\textwidth]{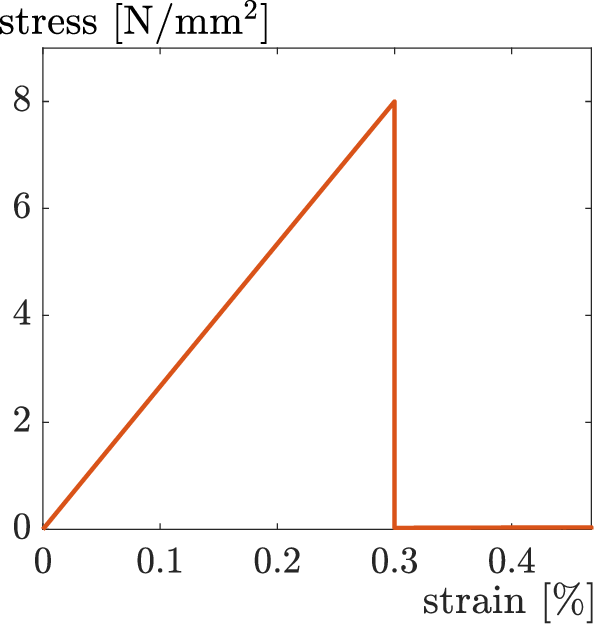}
		\caption{}
	\end{subfigure}
	
	\begin{subfigure}{0.3\textwidth}
		\includegraphics[width=\textwidth]{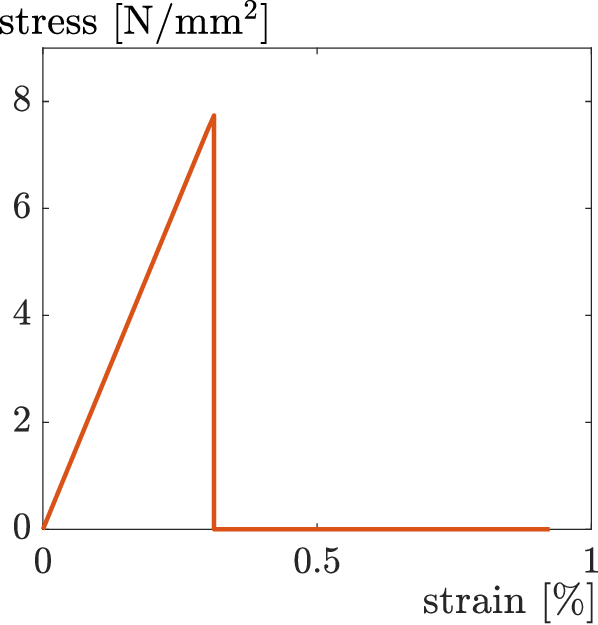}
		\caption{}
	\end{subfigure}
	\hfill
	\begin{subfigure}{0.3\textwidth}
		\includegraphics[width=\textwidth]{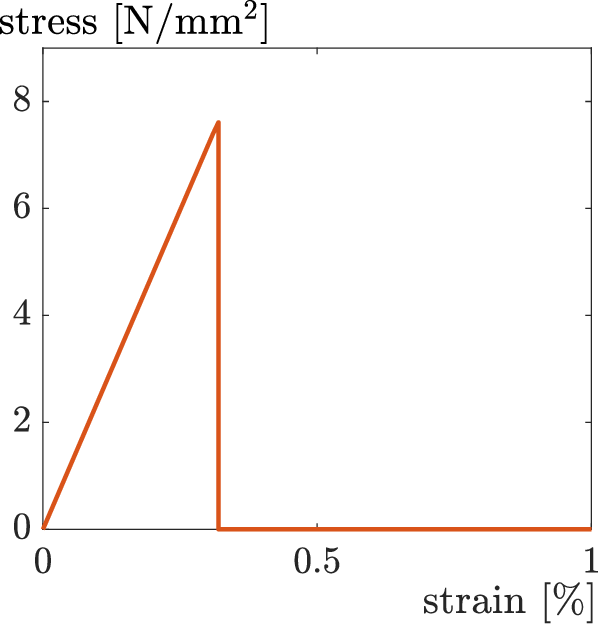}
		\caption{}
	\end{subfigure}
	
	\caption{stress-strain-diagrams for the simulations for (a) $0^{\circ}$, (b) $30^{\circ}$, (c) $45^{\circ}$, (d) $60^{\circ}$ and (e) $90^{\circ}$ using the ED-technique} \label{Plot_ED}
	
\end{figure}

Simulations were carried out using both the EE technique and the ED technique, and the results are discussed in the following. The failure patterns and the corresponding stress-strain-diagram obtained with EE simulations are shown in figures \ref{Bruchbilder_EE} and \ref{Plot_EE}, those obtained with ED in figures \ref{Bruchbilder_ED} and \ref{Plot_ED}.

In general, both methods lead to very similar results, which are also consistent with the experimental findings from \cite{Kiendl.2020}, in the sense that fracture always occurs in the intralayer zones except for the $0^\circ$ case. \textcolor{Rev}{The only difference is that the failure "jumps" from one intralayer zone to a neighbouring intralayer zone in the experiments. This could be related to the turnarounds or pre-damage from the printing process. For global failure to occur for the specimen, it seems that the "jump" from two neighbouring intralayer zones needs less energy then failure through the turnaround at the specimen's edge.  Additionally \cite{Kiendl.2020} shows that the maximum stress decreases with increasing angle between strand orientation and loading direction, which can be simulated qualitatively as well with the presented approach.} 
However, only with ED, it is possible to simulate all experiments until complete failure, while with EE, the simulations are often terminated before final failure of the specimens.
This can be seen most clearly for the $0^\circ$ case. Figure \ref{Bruchbilder_EE}(a) shows the crack pattern from the last successful simulation step. It can be seen, that the specimen is still far away from being completely broken, there are two cracks covering only approximately half of the specimen's width. It can also be seen in figure \ref{Plot_EE}(a), where the stress-strain curve stops shortly after reaching the maximum stress, still representing a stress value of approx. $35$ $MPa$. The same problem can be seen in figures \ref{Bruchbilder_EE}(c) and (e). In these cases, two cracks have developed, but none of them is going completely through the specimen, and the upper and lower parts of the specimen are still connected. This is confirmed by the stress-strain curves in figure \ref{Plot_EE}(c) and (e), where the curves stop shortly before reaching $\sigma=0$. Only for the $30^\circ$ and $60^\circ$ cases, the simulations can run until complete failure, which can be seen in figure \ref{Bruchbilder_EE}(b) and (d), where the specimen is clearly broken in two parts, and in the curves in \ref{Plot_EE}(b) and (d), dropping down to $\sigma=0$.\footnote{It's not completely zero due the very low stiffness in the eroded elements.} In all the cases, which do not reach the final failure, the simulations are terminated by the FE software due to excessively distorted elements within the fractured zone. Unfortunately, there is no option to ignore these elements in the distortion check. 
With ED instead the elements in the fractured zone are removed from the mesh and have no effect on the further simulation. As a consequence the simulations can run until complete failure of the specimens for all cases, as can be seen in figures \ref{Bruchbilder_ED} and \ref{Plot_ED}. \textcolor{Rev}{The global failure only appears between one intralayer zone, but there are many different areas of different intralayer zones where local failure occurs. This can be seen, for example, in the failure patterns using the EE technique in figure \ref{Bruchbilder_EE}, as well as in Digital Image Correlation images as showed in \cite{Cunha.2021}. After failure occurs these areas are difficult to recognize visually without any further post-processing using the ED technique due to a very small element size.}
From this study we can conclude, that the ED technique is more robust and better suited for these kind of simulations than EE, if the goal is to simulate the specimen's behaviour until complete failure.

\subsection{von Mises equivalent stress vs. Rankine equivalent stress}\label{unidirectional}
\begin{figure}[h]
	\centering
	
	\begin{subfigure}[][][c]{0.48\textwidth}
		\includegraphics[width=\textwidth]{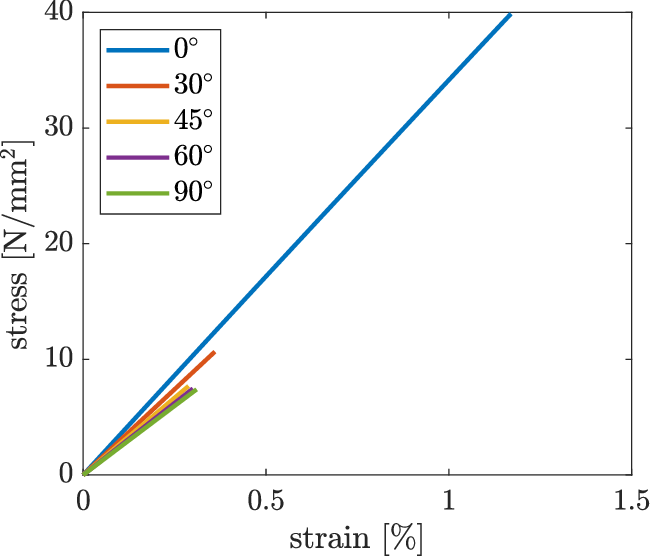}
		\caption{}
	\end{subfigure}
	\hfill
	\begin{subfigure}[][][c]{0.48\textwidth}
		\includegraphics[width=\textwidth]{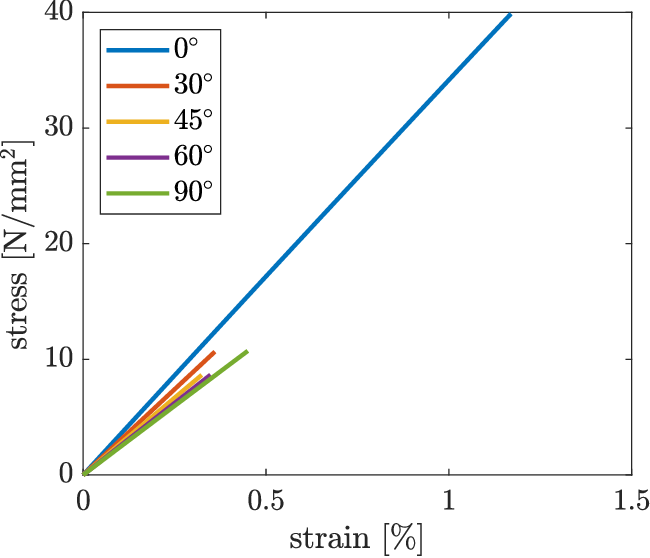}
		\caption{}
	\end{subfigure}	
	
	\caption{Stress-strain-diagrams using (a) Rankine equivalent stress (b) von Mises equivalent stress} \label{simulation_uni}
	
\end{figure}

In this section a numerical study comparing two different equivalent stress measures as failure criteria, namely the von Mises (equation \ref*{vonMises}) and Rankine (equation \ref*{Rankine}) equivalent stresses, is performed in order to find out which one is best suited to replicate the failure behaviour of FFF specimens observed in \cite{Kiendl.2020}. 
Figure \ref{simulation_uni} shows the stress-strain curves for all specimens ($0^\circ, 30^\circ, 45^\circ, 60^\circ, 90^\circ$) obtained with (a) Rankine equivalent stress and (b) von Mises equivalent stress. It is noted, that in both cases the curves do not perfectly match with the experimental curves in \cite{Kiendl.2020}. In general, it can be seen that the experiment from figure \ref{experiments_chao} does not have a linear curve, but a slightly curved curve in the stress-strain diagram. However, with the Rankine equivalent stress as failure criterion, the results correctly represent the general trend that the maximum stress decreases with increasing angle between loading direction and strand orientation. Instead with the von Mises failure criterion the $90^\circ$ specimen exhibits a higher strength than the $60^\circ$ and $45^\circ$ cases, which is in contrast to the experiment in \cite{Kiendl.2020}.

\begin{table}
	\begin{tabular}{l | c c c c c}
			& $0^{\circ}$ & $30^{\circ}$ & $45^{\circ}$ & $60^{\circ}$ & $90^{\circ}$ \\
		\hline
		$E_{exp}$ & 3434 & 2762 & 2627 & 2334 & 2337 \\
		$E_{sim}$ (von Mises) & 3439 & 2962 & 2667 & 2488 & 2385 \\
		$E_{sim}$ (Rankine) & 3439 & 2962 & 2667 & 2488 & 2385  \\
		$\sigma_{f, exp}$ & 39.95 & 27.47 & 20.16 & 14.17 & 7.24 \\
		$\sigma_{f, sim}$ (von Mises) & 39.88 & 10.65 & 8.63 & 8.65 & 10.72 \\
		$\sigma_{f, sim}$ (Rankine) & 39.88 & 10.65 & 7.68 & 7.46 & 7.39 \\
	\end{tabular}
	\caption{Material parameter $E$ [MPa] and $\sigma_{f}$ [MPa] of the experiments and simulations}\label{Material_parameter_simulation}
\end{table}

The brittle behaviour at failure of the experiments \cite{Kiendl.2020} and the decreasing of the material parameters $E$ and $\sigma_{f,sim}$ are better simulated by the Rankine hypothesis.

\subsection{Modelling of the specimen's edge}\label{edge}
\begin{figure}[h]
	\centering
	
	\begin{subfigure}[][][c]{0.2\textwidth}
		\includegraphics[width=\textwidth]{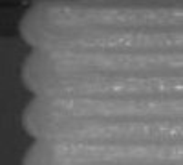}
		\caption{}
	\end{subfigure}
	\hfill
	\begin{subfigure}[][][c]{0.2\textwidth}
		\includegraphics[width=\textwidth]{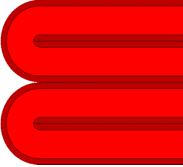}
		\caption{}
	\end{subfigure}
	\hfill
	\begin{subfigure}[][][c]{0.2\textwidth}
		\includegraphics[width=\textwidth]{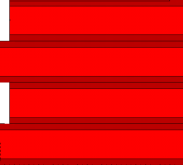}
		\caption{}
	\end{subfigure}	
	\hfill
	\begin{subfigure}[][][c]{0.2\textwidth}
		\includegraphics[width=\textwidth]{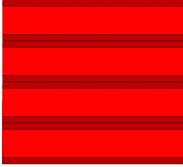}
		\caption{}
	\end{subfigure}		
	
	\caption{Edge of the $90^{\circ}$ specimen for (a) the experiment, (b) the simulation with turnarounds, (c) the simulation with staggered strands by two elements and (d) the simulation with non-staggered strands without turnarounds} \label{Probenrand}
	
\end{figure}

Since the FE-models represent the mesostructure, the question arises as to whether the edges have an influence on the simulations, as edges in the printed specimen are usually the initiation for a crack. Typically, the printing in the FFF results in connecting printed material between two strands, even if retraction is set in the slicer. These connections between the two strands cannot be avoided and could therefore potentially have an influence on the failure. 
A microscope image of the edges of  a $90^{\circ}$ specimen is depicted in figure \ref{Probenrand}(a), showing the typical "turnarounds" which are the result of the continuous print path. Since the geometry creation and meshing of these turnarounds is not trivial, especially for the non-$90^{\circ}$ cases, different ways of modelling with different level of simplification as shown in figure \ref{Probenrand}(b)-(d) are tested in the following. Figure \ref{Probenrand}(b) represents the reference model, where the turnarounds are modelled in detail. Figure \ref{Probenrand}(c) is a simplified model without turnarounds, but at the edge is always a slight offset between two neighbouring strands. The idea of this offset is to have reentrant corners, which create stress concentrations, that might have a similar effect as the reentrant corner between two turnarounds. Lastly,  figure \ref{Probenrand}(d) shows the most simplified model without turnarounds and a smooth edge. 

\begin{figure}[h]
	\centering
	\includegraphics[width=0.5\textwidth]{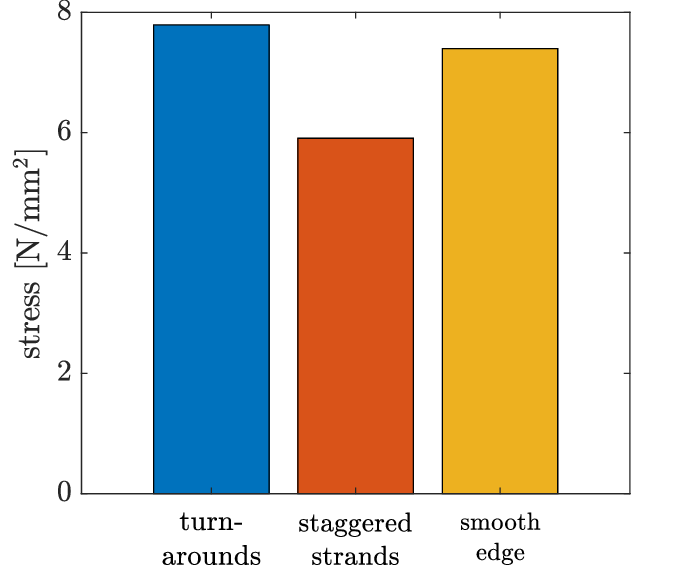}
	\caption{Maximum stress for the different types of modelling the edges of the $90^{\circ}$ specimen}\label{Simulation_Probenrand}
\end{figure}

Figure \ref{Simulation_Probenrand} shows the maximum stress obtained with the different types of modelling the edges. Noteworthy, the additional material at the edge (connecting strands/turnrounds) has only a minor influence of $5$ $\%$ on $\sigma_{f,sim}$ compared to the specimen without connecting strands. The staggered strands have a $25$ $\%$ lower $\sigma_{f,sim}$ and therefore a too big influence compared to the connecting strands. Therefore, due to the simpler modelling, it is recommended to carry out the simulations with an edge without turnarounds.

\subsection{Intralayer zone}\label{intralayer_zone}
\begin{figure}[h]
	\centering
	\includegraphics[width=0.6\textwidth]{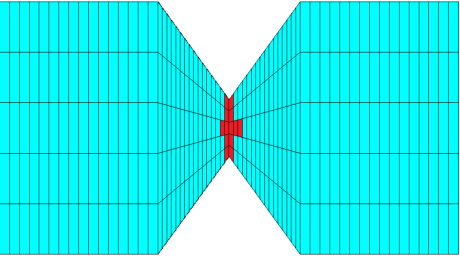}
	\caption{Side view of the intralayer zone of a  $90^{\circ}$ specimen. Deleted elements after failure are colored in red.}\label{damage_strand_ED}
\end{figure}

In section \ref{Mesh} the importance of the fineness of the mesh, especially in the intralayer zone, is described. Since the intralayer zone is responsible for the failure in all cases except for the $0^{\circ}$ specimen, it is important that only very few elements are deleted upon failure due to fine meshing. As shown in figure \ref{damage_strand_ED} only one and a maximum of up to three elements are deleted per cross-section of one strand, which corresponds to only $1.7$ $\%$ and $5$ $\%$  respectively of the strand's width $a$. 

\begin{figure}[h]
	\centering
	\includegraphics[width=0.5\textwidth]{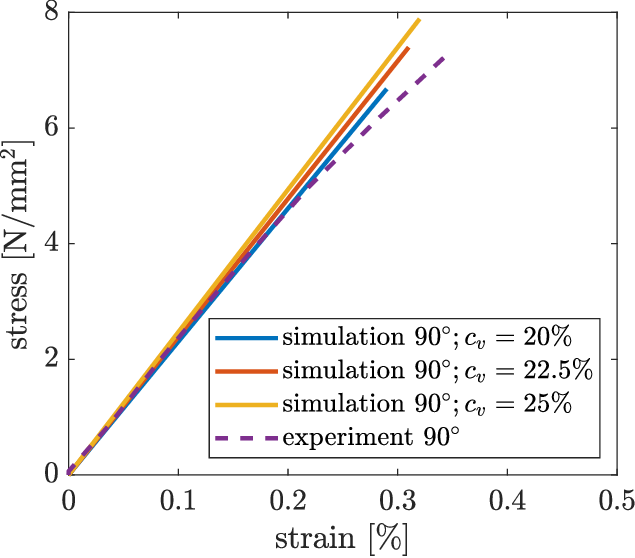}
	\caption{Stress-strain-diagrams for a $90^{\circ}$ specimen with various $c_v$}\label{Kontaktlaenge}
\end{figure}

As described in chapter \ref{geometry}, the vertical contact length of the individual strands in the intralayer zone has to be verified. The $90^{\circ}$ specimen is used for this, as no overlapping failure mechanisms other than the failure of the intralayer zone orthogonal to the strand are to be expected here. To ensure that the area of the strand remains constant, the horizontal contact length $c_h$ must also be changed when the vertical contact length $c_v$ is changed. The constant area is necessary so that the strength $\sigma_{f}$ and Young's modulus $E$ from table \ref{material_parameter} do not change for the $0^{\circ}$ specimen. For a $90^{\circ}$ specimen figure \ref{Kontaktlaenge} shows that $E$ and $\sigma_{f}$ also increase with increasing $c_v$. The best agreement between experiment and simulation is achieved with a $c_v = 0.225$. 

\section{Conclusion}
We have presented a finite element modelling approach for simulating the mechanical behaviour of unidirectional FFF-printed PLA specimens under tensile loading with different strand orientations. The goal was to represent the main features, which were observed experimentally in \cite{Kiendl.2020}. In contrast to other existing works, the focus in this paper is on the fracture behaviour, and the goal was to simulate the experiments until final failure of the specimens. In our approach, the mesostructure is resolved by using octagon cross-sections for each strand. We have shown that with such models, we can replicate the typical behaviour of FFF specimens in which failure typically happens between strands unless the strands are \textcolor{Rev}{almost} parallel to the loading direction and the specimens' strength decrease with increasing angle between loading direction and strand orientation. We have investigated several detail questions for the modelling, for example, we compared the suitability of Element Erosion and Element Deletion techniques for this kind of simulations, finding that only with Element Deletion, it was possible to simulate the loading until final failure for all cases. Furthermore, we compared different failure criteria, namely the von Mises and Rankine equivalent stresses, finding that only with Rankine equivalent stress, the simulation results show \textcolor{Rev}{a} typical behaviour\textcolor{Rev}{, which is in accordance to experiments like those in \cite{Kiendl.2020}, \cite{Yao.2020} or \cite{Khosravani.2022}}, of monotonic decrease in strength with increasing strand orientation angle. 
We have also shown that the turnarounds at the edges of FFF-printed parts can be neglected in the simulation models, which highly simplifies the model creation and meshing. 
These simulations form the basis for improvements such as a more complex material model for further research. \textcolor{Rev}{Our approach can also be used to investigate the failure behaviour of more complex structures.} A further step of the numerical model is the extension to cross layups and the verification of the increase in toughness as shown experimentally in \cite{Kiendl.2020}. \textcolor{Rev}{Extending this approach to other filament materials, which require more complex material models like plasticity, hyperelasticity and viscoelasticity, is planned as future research. Additionally real world problems such as load capacity, failure behaviour and location for complex FFF printed parts due to non-uniform layer orientations can be investigated with the presented approach as well.}

\section{Acknowledgements}
This work has received funding from the European Research Council (ERC) under the European Union's Horizon 2020 research and innovation programme (grant agreement No 864482) and dtec.bw - Digitalization and Technology Research Center of the Bundeswehr. dtec.bw is funded by the European Union NextGenerationEU fund. The financial support by University of the Bundeswehr Munich enabled open access publishing. On behalf of all authors, the corresponding author states that there is no conflict of interest. Images used courtesy of ANSYS, Inc.
All the support is gratefully acknowledged.


\bibliography{FFF-Simulation_Maj_Rev}

\end{document}